\documentclass[preprint,12pt,longtitle,authoryear]{elsarticle}

\usepackage[margin=1.0in]{geometry} 
\usepackage{multicol}
\usepackage{subfigure}
\usepackage{epsfig}
\usepackage{latexsym}
\usepackage{epic}
\usepackage[leqno]{amsmath}
\usepackage{amssymb}

\newtheorem{theorem}{Theorem}

\def\O{\mathcal{O}}

\begin{document}
\begin{frontmatter}

\title{ Hide and seek: placing and finding an optimal tree for thousands of homoplasy-rich sequences}

\author{Dietrich Radel$^1$, Andreas Sand$^{2,3}$, and Mike Steel$^{1,*}$}
\address{$^1$Biomathematics Research Centre, University of Canterbury, Christchurch, New Zealand;\\ 
$^2$Bioinformatics Research Centre, Aarhus University, Denmark\\
$^3$Department of Computer Science, Aarhus University, Denmark\\
$^*$Corresponding author}

\vspace{1in}


\begin{abstract} Finding optimal evolutionary trees from
sequence data is typically an intractable problem, and there is
usually no way of knowing how close to optimal the best tree from some
search truly is. The problem would seem to be particularly acute when
we have many taxa and when that data has high levels of homoplasy, in
which the individual characters require many changes to fit on the
best tree. However, a recent mathematical result has provided a
precise tool to generate a short number of high-homoplasy characters
for any given tree, so that this tree is provably the optimal tree
under the maximum parsimony criterion. This provides, for the first
time, a rigorous way to test tree search algorithms on homoplasy-rich
data, where we know in advance what the `best' tree is.  In this short
note we consider just one search program (TNT) but show that it is
able to locate the globally optimal tree correctly for 32,768 taxa,
even though the characters in the dataset requires, on average, 1148
state-changes each to fit on this tree, and the number of characters
is only 57. \\

\end{abstract}
\end{frontmatter}

Phylogenetic tree reconstruction methods based on optimization
criteria (such as maximum parsimony or maximum likelihood) have long
been known to be computationally intractable (NP-hard)
\citep{Foulds1982}. However, on perfectly tree-like data (i.e. long
sequences with low homoplasy), these methods will generally find the
optimal tree quickly, even for large datasets.  Moreover, when data
is largely tree-like, there are good theoretical and computational
methods for finding an optimal tree under methods such as maximum
parsimony, with an early result more than 30 years ago
\citep{Hendy1980}, along with more recent developments
\citep{Blelloch2006, Holland2005}.

So far, it has not been clear whether such methods would be able to
find the global `optimal' tree for homoplasy-rich datasets with
large numbers of taxa, particularly when the sequences are short.  The
traditional view \citep{Sokal1963} is that homoplasy tends to obscure
tree signal, requiring more character data than homoplasy-free data to
recover a tree, though contrary opinions that homoplasy can `help'
have also appeared \citep{Kalersjo1999}.

A fundamental obstacle arises in trying to answer this question: One
usually cannot guarantee in advance that any tree will be optimal for
homoplasy-rich data without first searching exhaustively through tree
space, and this precludes datasets involving hundreds (let alone
thousands) of taxa. However, a recent mathematical result by
\cite{Chai2011} can be used to construct large synthetic datasets on
many (thousands) of taxa that simultaneously (i) have a high degree of
homoplasy, (ii) come with a guaranteed certificate as to what the
optimal tree will be under minimum evolution (maximum parsimony), and
(iii) have sequence lengths that are much shorter than the number of taxa.

We can thus, for the first time, test existing programs to see how
they perform in such settings, as we know ahead of the analysis what
the unique optimal tree is.  It might be expected that, with many taxa and high homoplasy,
finding this uniquely optimal tree would be impossible.
However, we show that this is not the case.
In particular, one program (TNT) is
able to correctly identify the uniquely most parsimonious tree on
thousands of taxa, each requiring many changes. In one 
case, the uniquely most parsimonious tree for 32,768 taxa was
successfully found, even though the characters in this dataset required, on
average, 1148 state-changes each to fit on this tree, and the
number of characters was only 57.  This search involved more than $6.1
\times 10^{13}$ tree rearrangements, but was completed in $12$ hours
on a common multi-purpose computer.

 Our results provide a positive message for molecular phylogenetics on two fronts:  (i) globally optimal
trees on thousands of taxa can be recovered from short sequences by existing software in reasonable time, and (ii) high levels of homoplasy, rather than erasing phylogenetic signal, can enhance it in certain settings, in line with  \cite{Kalersjo1999}.

\subsection*{Notation}
Throughout, we will let $n$ denote the number of taxa, and $k$ the
number of characters.  For a sequence $D= (c_1, c_2, c_3, \ldots,c_k)$ of characters, let
$s(c_i, T)$ denote the \emph{parsimony score} of $c_i$ on an $X$-tree $T$,
and let  $h(c_i, T)$ the \emph{homoplasy score} of $c_i$ on $T$
(see~\cite{Semple2003} for details).  In the case of binary characters
(excluding the constant character that assigns all taxa the same
state), $h(c_i, T) = s(c_i, T) - 1$.

Let $S(D, T) = \sum_{i=1}^k s(c_i, T)$ denote the parsimony score of
$D$ on $T$ and $H(D,T) = \sum_{i=1}^k h(c_i, T)$ denote the homoplasy
score of $D$ on $T$.  Thus if all the characters in $D$ are binary
(and not constant) then $H(D,T) = S(D,T) - k$.  Finally, let $H(D) =
\min_{T} H(D,T)$ and $S(D) = \min_T S(D, T)$ be the homoplasy score
and the parsimony score of the most parsimonious tree, respectively.

\subsection*{Consistency and retention indices} 
The consistency and retention indices have traditionally been used to
measure the amount of homoplasy in a set of characters $D$. The {\bf
  consistency index} $CI$ of a set of characters is
defined~\citep{Wiley2011} as the ratio $M / S$ where $M = \sum_{i =
  1}^k \min_T s(c_i, T)$ is the sum of the minimum number of steps for
each character and $S$ is the sum of the actual number of steps. A set of
characters with a consistency index of $1$ exhibits no homoplasy, and
the consistency index decreases as the amount of homoplasy
increases. In the case of binary characters, $M = k$ and $S$ is the
best score, and therefore $CI = k / S(D)$. The consistency index unfortunately
grows with the number of taxa, making it hard to compare $CI$ values
across datasets. To overcome this, the retention index can be
used. The {\bf retention index} ($RI$) is defined~\citep{Wiley2011} as
the ratio $( G - S ) / ( G - M )$ where $G = \sum_{i = 1}^k \max_T
s(c_i, T)$ is the sum of the maximum number of changes for each
character $c_i$ on any tree. For binary characters, $G$ corresponds to
the sum over all $c_i$ of the size of the smaller portion of the
two-partition of taxa determined by $c_i$.

\section*{Short binary sequences that have a uniquely most parsimonious
  tree are `noisy'}
It was recently shown \citep{Huber2005} that for any binary tree $T$
(with any number of leaves), there is a sequence $D$ of just
\emph{four} multi-state characters for which $T$ is the uniquely
most parsimonious tree.  Moreover, in that setting one also has $H(D,T) =0$;   that is,
the characters exhibit no homoplasy on $T$.  When we move to binary
characters, however, the situation is very different, as the next
result shows.

\begin{theorem}
  For any sequence $D$ of $k$ binary characters that has a uniquely
  most parsimonious tree on $n$ leaves,  we have:
  \begin{equation}
    H(D) \geq 2n-3-k.
  \end{equation}
  Furthermore, when $k=\O(\log n)$, the average homoplasy score per
  character on $T$ tends to infinity as $n$ grows.
  \label{theorem}
\end{theorem}

{\em Proof:} 
  $T$ is a \emph{uniquely} most parsimonious tree for $D$ if it is the
  only $X$-tree that realizes the minimal parsimony score for $D$.
  This implies that collapsing any edge of $T$ leads to a tree that is
  not most parsimonious for $D$. In particular,  if $T$ is a uniquely most
  parsimonious tree, then it must be a binary phylogenetic tree and so
  have exactly $2n - 3$ edges.

\begin{figure*}
    \begin{center}
      \resizebox{12cm}{!}{
        \includegraphics{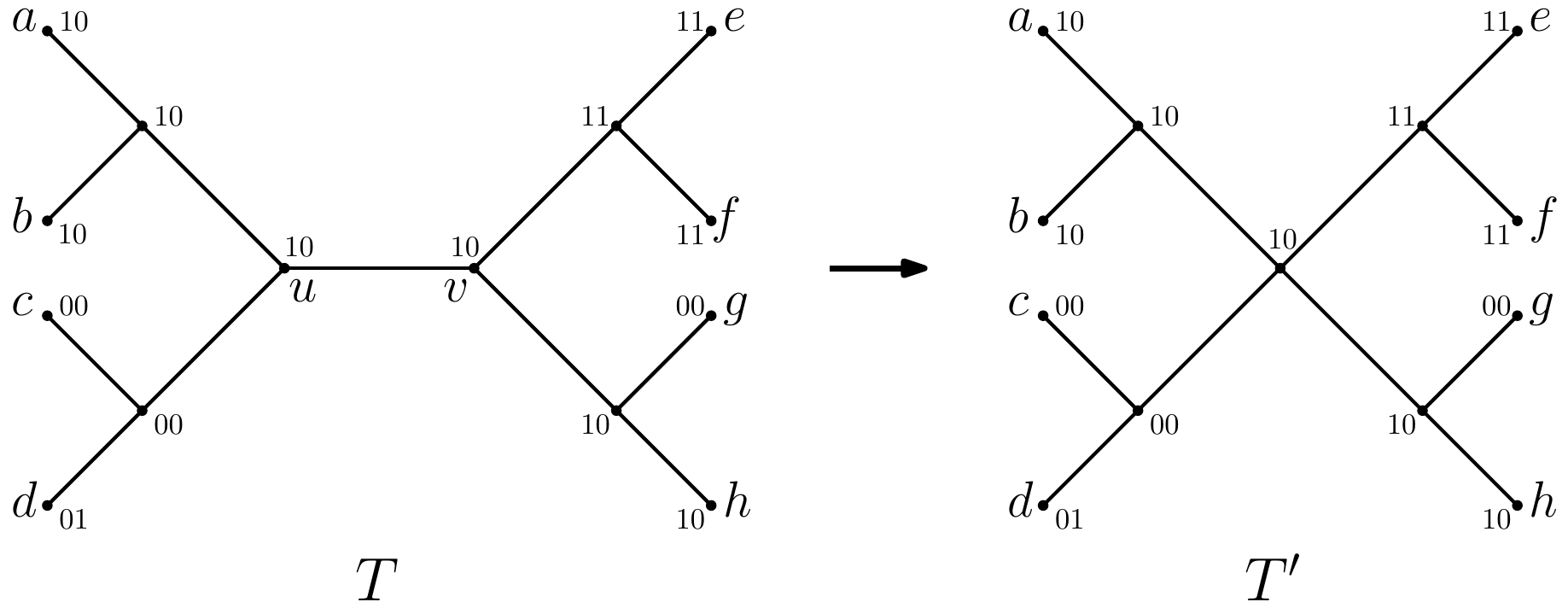}
      }
      \caption{$T' = T \backslash e$; removing an edge with no change
        for any character does not change the parsimony score.}
      \label{figure3}
    \end{center}
  \end{figure*}

  Let $T$ be the uniquely most parsimonious tree for $D = \{c_1,
  c_2,.. ..,c_k\}$, and fix a most parsimonious reconstruction
  $\overline{c_i}$ of each character $c_i$ on $T$ (thus
  $\overline{c_i}$ is an assignment of states to the vertices of $T$
  that extends the leaf assignment $c_i$).  Suppose that $T$ has an
  edge $e = \{u, v\}$ for which the $\overline{c_i}(u) =
  \overline{c_i}(v)$ for all $i \in \{1, \ldots, k\}$.  Then for the
  tree $T'=T \backslash e$ obtained from $T$ by collapsing the edge $e =
  (u,v)$ (a simple example, where $k=2$ is shown in
  Fig.~\ref{figure3}), we have: $S(D,T') = S(D,T)$, which is a
  contradiction, since $T$ is assumed to be the {\em only} most
  parsimonious tree for $D$.  So, for every edge of $T$, we must have
  $\overline{c_i}(u) \neq \overline{c_i}(v)$ for at least one $i \in
  \{1, \ldots, k\}$.  For an edge $e = \{u,v\}$ of $T$ and a character
  $c_i$ of $D$, let:
  \begin{equation}
    I(e,c_i)=
    \left \{
      \begin{array}{l l}
        1, \text{if $\overline{c_i}(u) \neq \overline{c_i}(v)$;}\\
        0, \text{otherwise.}
      \end{array}
    \right.
  \end{equation}

  \noindent Then, from our argument above, $\sum_{i=1}^k I(e,c_i) \geq
  1$ for each edge $e$ in $T$ and $s(c_i, T) = \sum_{e} I(e,c_i)$
  for every $c_i$ in $D$. Thus we have:
  \begin{equation}
    \begin{aligned}
      S(D, T) &= \sum_{i=1}^k s(c_i, T) = \sum_{i=1}^k \sum_{e} I(e,c_i)\\
      &= \sum_{e} \sum_{i=1}^k I(e, c_i) \geq 2n-3,  \\
    \end{aligned}
  \end{equation}
  as $T$ has $2n - 3$ edges. Hence, since every $c_i$ is binary and $T$
  is the most parsimonious tree, we get:
  \begin{equation}
    H(D) = S(D) - k \geq 2n-3-k.
  \end{equation}
  For the second claim, note that:
  \begin{equation}
    \frac{H(D)}{k} \geq \frac{2n - 3}{k} - 1 \rightarrow \infty \text{ as } n \rightarrow \infty,
  \end{equation}
  when $k = \O(\log n)$.
\hfill$\Box$

\section*{An explicit construction}
\begin{figure}
  \centering 
  \subfigure[]{
    \includegraphics[width=0.5\columnwidth]{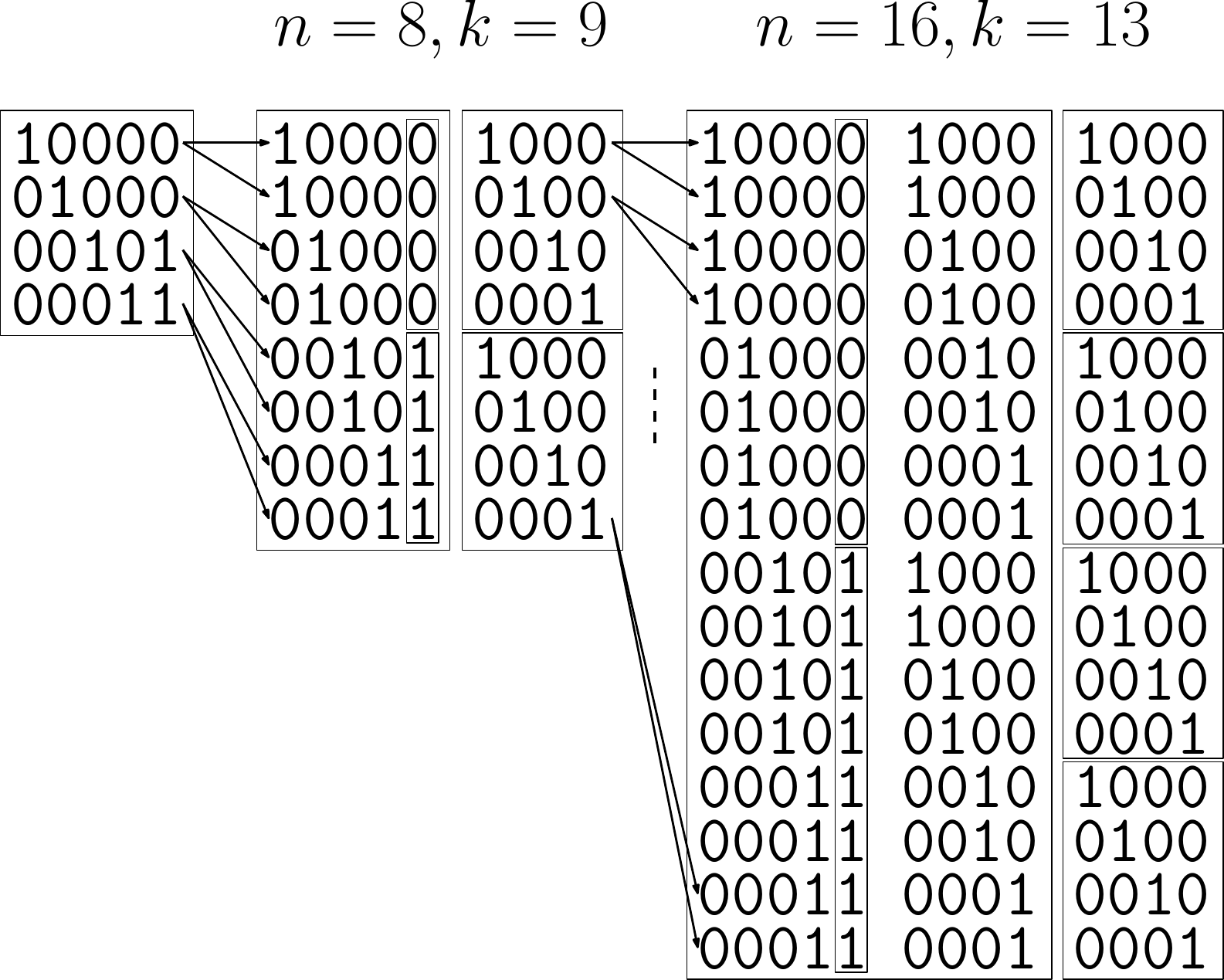}
    \label{fig:algorithm}
  }\\
  \subfigure[]{
    \includegraphics[width=0.5\columnwidth]{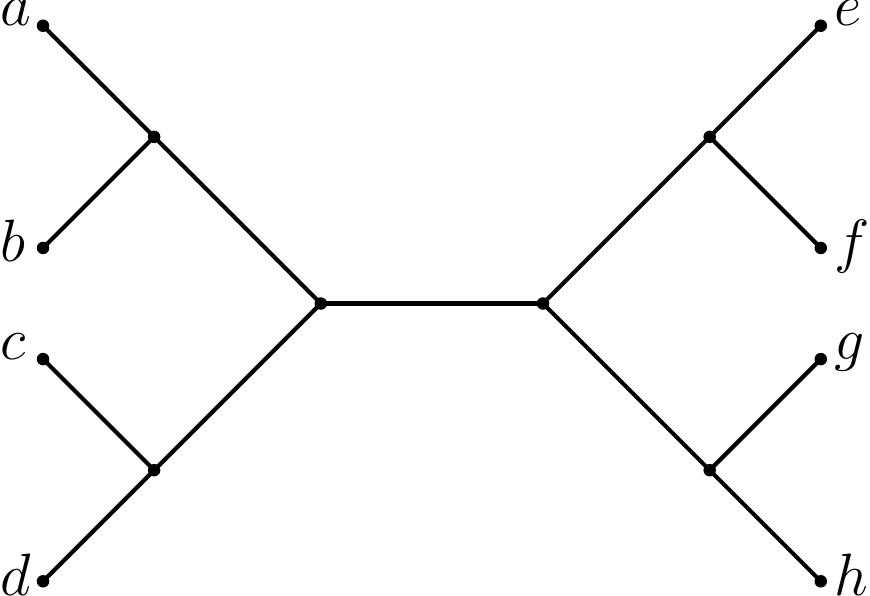}
    \label{fig:8_chars}
  }
  \caption{The balanced tree construction based on
    ~\cite{Chai2011}. (a) The number of taxa can be doubled by repeating each taxon and then adding  four
    new characters at the end of each sequence, repeating the pattern
    $1000$, $0100$, $0010$, $0001$. (b) The tree topology for
   eight taxa, where $a, b, \ldots, h$ are labeled with the $1$st,
    $2$nd, $\ldots 8$th row in column two of (a). The interior nodes
    can be labeled such that each edge has a change for exactly one
    character, and the parsimony score therefore equals the number of
    edges.}
  \label{fig:construction}
\end{figure}

We applied the construction described in~\cite{Chai2011}, which allows
the construction, for each integer value of $p\geq 2$, of a set $D$ of $n=2^p$ taxa sequences with $k=4p -
3$ characters each and with a uniquely most parsimonious tree with the
parsimony score $S(D)= 2n - 3$  (see
Fig.~\ref{fig:construction}). The consistency index for all these datasets is given by:
$CI = \frac{k}{S(D)} =\frac{4p-3}{2^{p+1} - 3}$, which converges
exponentially fast to $0$ when $p$ (and thereby the number of taxa)
grows towards infinity. Moreover, for calculating the retention index
(RI) we have: $G$= $\frac{n}{4}(k - 1) + \frac {n}{2} = 2^p (p -
\frac{1}{2})$. To see this, recall that $G$ is the sum over all
characters $c_i$ of the size of the smaller portion of the partition of
taxa determined by $c_i$, and note that in the construction
illustrated in Fig.~\ref{fig:algorithm} the last character of the
first block always contains $n/2$ taxa in state $1$ and the remaining
characters have $n/4$ taxa in state $1$. For the homoplasy per
character,$h(D)$, we have $h(D) = H(D)/k = \frac{2n-3-k}{k} =
\frac{2^{p+1} - 3}{4p - 3} - 1$, which tends to infinity as $p$ (and
thereby the number of taxa) goes to infinity (as stated in
Theorem~\ref{theorem}). We note that this matches the bound given in
Theorem~\ref{theorem}, and thus shows that this bound can be realized.

\section*{Results}
To test how well TNT~\citep{Goloboff2008} could recover phylogenies
with high amounts of homoplasy, we generated datasets with $n = 8, 16,
32, \ldots 32,768$ taxa according to the construction
by~\cite{Chai2011}, and ran TNT version 1.1 64 bit (May 2012, Linux 64 version) to
recover the phylogenies by maximum parsimony. The search heuristics of
TNT can be guided by the user, based on knowledge about the input by
setting multiple command options. But as we were mainly interested in
the default performance, we did not use these. However, to speed up
the computations, we used the option ``\texttt{xmult=level x;}''
(where \texttt{x} is a number between $0$ and $10$). The authors of
TNT recommend using level $0$--$2$ for easy datasets, $3$--$5$ for
medium, and $6$--$10$ for hard datasets.  These experiments are
summarized in Table~\ref{tab:results}. Each experiment was performed
in a single thread on a dual Intel\textsuperscript{\textregistered}
Xeon\textsuperscript{\textregistered} CPU (3.07GHz, six cores each)
computer running openSUSE v. 11.x.

In Table~\ref{tab:results} first of all note that the most
parsimonious tree was successfully reconstructed in all
experiments. It was found almost instantly for datasets with up to $n
= 128$ taxa (using ``\texttt{xmult=level 1;}''). For datasets with up
to $n = 16,384$ taxa, the most parsimonious trees were found within
two hours (using ``\texttt{xmult=level 3;}''). And even for the
dataset with $n = 32,768$ taxa, the most parsimonious tree topology
was identified, although this computation took approximately twelve
hours (using ``\texttt{xmult=level 4;}'' which is used for medium-difficulty datasets).

In another set of experiments (results not shown) we tested TNT on
datasets which were built using another construction
by~\citeauthor{Chai2011} such that the uniquely most parsimonious tree
for each dataset is a caterpillar tree. Thus these experiments tested
TNT's performance on the opposite pole of tree space. The uniquely
most parsimonious trees were also successfully reconstructed across a selection of these
earlier analyses, although the reconstruction was significantly more
time consuming (e.g. 2 hours 32 minutes for $n = 8192$). This was
unexpected, and the reason for it is unclear to us, but it may be a
property of TNT's search heuristic.

\begin{table*}[ht]
  \centering
    \includegraphics[width=0.7\textwidth]{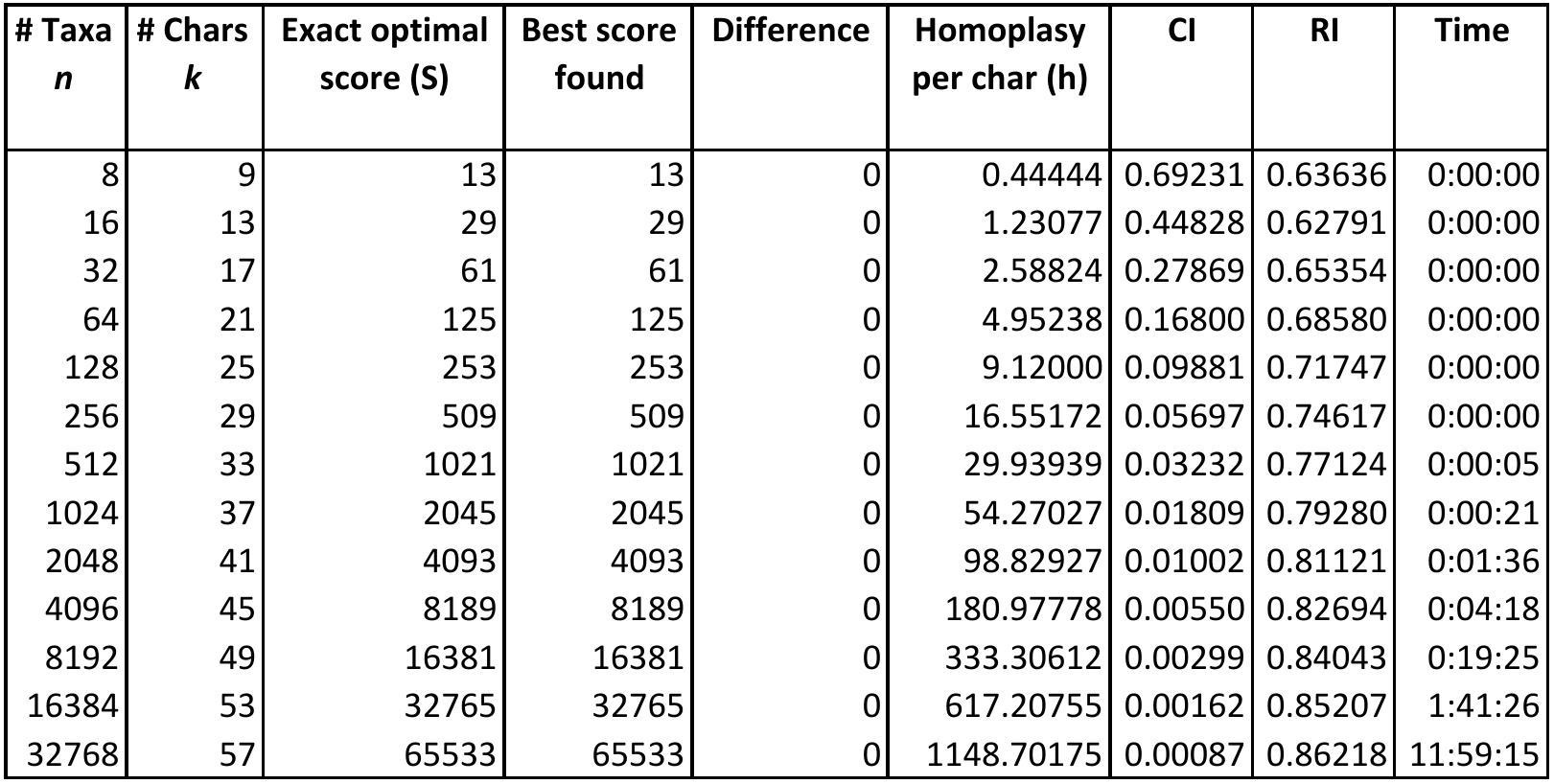}
  \caption{Summary of experiments on the balanced tree topology. The uniquely most parsimonious tree
    was successfully reconstructed in all experiments.}
  \label{tab:results}
\end{table*}

\section*{Concluding comments}
The mathematical foundation provided by the Chai--Housworth
construction opens the door to a unique experiment that has been
impossible until now:  searching for the most parsimonious tree in a
dataset involving sequences on large numbers of taxa and with high
homoplasy, for which we know in advance what the most parsimonious
tree is.  It was not at all clear whether existing phylogenetic
programs would be able to locate this most parsimonious tree in such a
large tree space (when $n = 2^{15}$, the search space contains more
than $10^{140000}$ trees), yet at least one program (TNT) was able to
do so.  It was not the intention of this short note to compare
different parsimony programs on this test dataset, but that would
surely be a reasonable project for future work.

Also, the question of whether the homoplasy present in this data is a
good proxy for `noise' in biological data is difficult to determine --
the construction by~\cite{Chai2011} does have an obvious pattern and
structure, so it might be argued that finding a tree for such data may
be inherently easier than for data for which the homoplasy comes about
through random processes.  However, there is currently no way to
guarantee what the maximum parsimony tree would be for random data,
though a conjecture (conjecture 1.3.1 in \cite{Albert2005}), if
established, would provide one.  Nevertheless, we find it surprising
that one can find a uniquely most parsimonious tree on more than
$3\times10^4$ taxa with just $57$ characters that require, on average,
more than $1000$ substitutions to fit on the best tree.

A further task that would be worthy of study would be to investigate
the influence of different tree shapes, beyond the symmetric branching
trees and the caterpillar trees considered here, using the general
construction given by~\citeauthor{Chai2011}. This is, however, again
beyond the scope of this short note.

\section*{Acknowledgments}  We thank the {\em Allan Wilson Centre for Molecular Ecology and Evolution} for
helping fund this research. We also acknowledge use of the program TNT, which is made available with the sponsorship of the Willi Hennig Society.

\bibliographystyle{elsarticle-harv}
\bibliography{note_MPE}

\end{document}